\documentclass[a4paper,11pt]{article}
\usepackage{jheppub} 
\usepackage{lineno}
\usepackage{amssymb}
\usepackage{bbold}
\usepackage[vietnamese,english]{babel}  
\usepackage{tikz}
\usetikzlibrary{shapes.geometric}
\usepackage[compat=1.1.0]{tikz-feynman} 
\usepackage{slashed}
\usepackage{mathtools}

\newcommand{\op}[2]{ \mathcal{O}_{#1} ^{#2} } 
\newcommand{\coeff}[2]{ c_{#1} ^{#2} } 

\usepackage[compat=1.1.0]{tikz-feynman}
\usepackage{subcaption}
\pgfmathsetmacro\sizedot{1.1}
\pgfmathsetmacro\sizesqdot{1.6}
\pgfmathsetmacro\sizeemdot{2.}

\DeclareFontFamily{OT1}{pzc}{}
\DeclareFontShape{OT1}{pzc}{m}{it}{<-> s * [1.10] pzcmi7t}{}
\DeclareMathAlphabet{\mathpzc}{OT1}{pzc}{m}{it}
 
\newcommand{\yuk}[1]{{y}_{#1}} 

\newcommand{\PR} {\mathbb{P}_R}
\newcommand{\PL} {\mathbb{P}_L}

\makeatletter
\newcommand{\overleftrightsmallarrow}{\mathpalette{\overarrowsmall@\leftrightarrowfill@}}
\newcommand{\overrightsmallarrow}{\mathpalette{\overarrowsmall@\rightarrowfill@}}
\newcommand{\overleftsmallarrow}{\mathpalette{\overarrowsmall@\leftarrowfill@}}
\newcommand{\overarrowsmall@}[3]{%
	\vbox{%
		\ialign{%
			##\crcr
			#1{\smaller@style{#2}}\crcr
			\noalign{\nointerlineskip}%
			$\m@th\hfil#2#3\hfil$\crcr
		}%
	}%
}
\def\smaller@style#1{%
	\ifx#1\displaystyle\scriptstyle\else
	\ifx#1\textstyle\scriptstyle\else
	\scriptscriptstyle
	\fi
	\fi
}
\makeatother

\pgfmathsetmacro\sizedot{1.1}
\pgfmathsetmacro\sizesqdot{1.5}
\pgfmathsetmacro\sizecrodot{1.0}

\title{Mapping between $\gamma_5$ schemes in the Standard Model Effective Field Theory}

\author[a]{S. Di Noi,}
\author[b]{R. Gr\"ober,}
\author[b]{P. Olgoso.}
\emailAdd{stefano.dinoi@kit.edu, ramona.groeber@pd.infn.it, pablo.olgosoruiz@unipd.it }
\affiliation[a]{Institute for Theoretical Physics, Karlsruhe Institute of Technology, 76131 Karlsruhe, Germany}
\affiliation[b]{Dipartimento di Fisica e Astronomia "Galileo Galilei", Universit\`a di Padova, Italy and Istituto Nazionale di Fisica Nucleare, Sezione di Padova, Padova, I-35131, Italy}

\abstract{We explore the relation between two distinct prescriptions for $\gamma_5$ in dimensional regularization  -- the Breitenlohner Maison t'Hooft Veltman (BMHV) scheme and Naive Dimensional Regularisation (NDR). The BMHV scheme is the only algebraically consistent scheme, but necessitates chiral symmetry restoring counterterms and it is computationally more expensive, limiting its practical use. We show how the quantum effective action can be translated between both schemes and present these translation rules for the Wilson Coefficients of the Standard Model Effective Field Theory (SMEFT), that can easily be implemented into automated tools for SMEFT computations.
Finally, we examine how this scheme dependence manifests itself in matching calculations, identifying the cases in which the dependence cancels in the final result.
To examplify this, we consider a concrete UV scenario matched onto the SMEFT at one-loop order in both schemes.
 Our work aims to facilitate more accurate SMEFT computations and can be considered as a first step towards a comprehensive map between the two continuation schemes. 
}
\preprint{KA-TP-10-2025}
\begin{document}

\maketitle
\flushbottom

\section{Introduction}\label{sec:intro}

The observation of a scalar particle compatible with the Higgs boson of the Standard Model (SM) in 2012 \cite{HiggsATLAS,HiggsCMS} has completed the mosaic of SM particles.
Several hints, such as a missing dark matter candidate or the baryon asymmetry of the universe point towards the necessity of extending the SM. 

The lack of direct evidence for physics beyond the Standard Model suggests that new physics may emerge at a scale higher than the electroweak scale. This motivates a general approach parametrizing new physics by means of an Effective Field Theory (EFT). A particularly popular approach is Standard Model Effective Field theory (SMEFT) \cite{Buchmuller:1985jz,dim6smeft}
used to to describe the effects of heavy new physics lying beyond the current experimental reach. These effects are parametrized by higher-dimensional operators that respect the symmetries of the SM. We will consider operators up to dimension six.

The increasing experimental precision requires more and more precise theory predictions, for which loop computations are often necessary. In this context, divergences arise; a regularisation scheme must be prescribed. The most common choice is represented by dimensional regularisation \cite{THOOFT1972189, Bollini:1972ui, Ashmore:1972uj, Cicuta:1972jf},\footnote{See Ref.~\cite{Gnendiger:2017pys} for other regularisation schemes.} where the number of space-time dimensions is promoted to a continuous value, $4 \to D=4-2\epsilon$. Subtleties arise in presence of chiral interactions, with the fifth element of Dirac algebra, $\gamma_5$, being an intrinsically four-dimensional object: a continuation scheme must be employed. The most common choices are \textit{na\"ive dimensional regularisation} (NDR) \cite{CHANOWITZ1979225} and the \textit{Breitenlohner-Maison-'t Hooft-Veltman} scheme (BMHV) \cite{THOOFT1972189,Breitenlohner:1977hr}, on which we will focus in the following.
The consistent continuation of $\gamma_5$ to $D$ dimensions becomes particularly pressing in the SMEFT, where a large number of operators containing chiral structures appear in loop computations. The recent years have seen a lot of advances both in methodology and in  precision results in multi-loop EFT computations \cite{Jenkins:2023bls,Fuentes-Martin:2023ljp,Fuentes-Martin:2024agf,Born:2024mgz,Naterop:2024ydo,DiNoi:2024ajj,Haisch:2024wnw,Duhr:2025zqw}, so the issue is becoming more and more important. 

The BMHV scheme is the only self-consistent scheme in which Ward identities have been proven to be valid to all orders in pertubation theory \cite{Speer:1974cz, Breitenlohner:1975hg, Breitenlohner:1976te}, relying on a rigorous mathematical definition of $\gamma_5$, while the NDR scheme becomes ambiguous in the presence of traces involving $\gamma_5$ and at least six $\gamma$ matrices. For this reason, the latter has been augmented with reading point descriptions \cite{Kreimer:1993bh, Korner:1991sx}, which though at three-loop order can become inconsistent and need to be extended \cite{Chen:2023ulo,Chen:2024zju}. More importantly, these prescriptions were not proven to work in general, so one needs to check on a case by case scenario, rendering it very inefficient. On the other hand, the BMHV scheme breaks chiral symmetries, that need to be restored by gauge non-invariant counterterms. Those have been computed in various chiral gauge theories in the recent years \cite{Belusca-Maito:2020ala, Belusca-Maito:2021lnk, Cornella:2022hkc, Belusca-Maito:2023wah}, but unfortunately they are still missing for the SMEFT, which makes the use of the BMHV scheme cumbersome. Apart from the necessity of counterterms, the BMHV scheme comes also with some practical issues: being algebraically more challenging it generates large expressions in intermediate steps of loop computations. For instance, it requires also extra steps when using the usual key-chain to perform loop computations at two-loop order or higher, i.e.~for instance when using integration by parts relations to reduce the loop integrals to a smaller set of master integrals \cite{Heller:2020owb}.

In this work, we do a first step towards a complete map between the NDR and BMHV scheme. This will aid to obtain results in the consistent BMHV scheme with tools that are unable to use it for practical reasons, setting the ground for higher order calculations. For the moment we restrict ourselves to the gauge interactions being QCD and therefore vector-like, i.e., the limit $g_1,g_2\rightarrow 0$, avoiding the need of the gauge symmetry-restoring counterterms. We pay though special attention to restore also the chiral global symmetry by following the formalism of \cite{OlgosoRuiz:2024dzq} such that our results provide a first step towards a map between the two continuation schemes for the full SMEFT. 

This paper is structured as follows: in Sec.~\ref{sec:framework} we explore the relation between the two schemes and present the framework we use for our translation.
Sec.~\ref{sec:examples} is devoted to show an explicit example of our calculation to further clarify some technical aspects.  In Sec.~\ref{sec:results} we describe the implementation of our framework in \texttt{Matchmakereft} \cite{Carmona:2021xtq} that we used to obtain our results.
Finally, in Sec.~\ref{sec:matching} we comment on scheme differences in matching calculations and how the scheme differences manifest in an explicit complete one-loop computation. In Sec.~\ref{sec:conclusion} we give our conclusions.

\section{Framework \label{sec:framework}}
In quantum field theory, loop-order calculations in perturbation theory often contain divergences. In order to remove these divergences from physical predictions, the theory must be renormalized, requiring a regularization scheme. 
The most common one is 
dimensional regularization, in which the action is extended to $D=4-2\epsilon$ space-time dimensions, with divergences regularized by poles in $1/\epsilon$ for $\epsilon\to 0$. In theories with fermions this requires the definition of a $D$-dimensional Dirac algebra.
The physical, four-dimensional properties must remain unchanged, so different choices involve vanishing quantities in the limit $D\to4$, referred to as evanescent objects, and simply define (as long as they are consistent) different schemes. 

The straightforward approach is to extend the same anticommutation relations to $D$-dimensions, i.e.:
\begin{equation}
    \{\gamma_\mu,\gamma_\nu\}=2\eta_{\mu\nu},\qquad  \{\gamma_\mu,\gamma_5\}=0.
\end{equation}
This strategy is called \textit{na\"ive dimensional regularization} (NDR). However, it is not possible to simultaneously keep the following four-dimensional properties:
\begin{itemize}
    \item[i)] $\{\gamma_\mu,\gamma_5\}=0$,
    \item[ii)] $\mathrm{Tr}[\gamma_\mu\gamma_\nu\gamma_\rho\gamma_\sigma\gamma_5]=-4\mathrm{i}\epsilon_{\mu\nu\rho\sigma}$,
    \item[iii)] Cyclicity of the trace.
\end{itemize}
In fact, traces of $\gamma_5$ with six or more $\gamma$ matrices are \textit{ambiguous}, yielding different results if one computes the trace from different starting points:
\begin{equation}
	\text{Tr}[\gamma_{\mu_1} \gamma_{\mu_2} \dots \gamma_{\mu_{2n}}\gamma_5] = \text{Tr}[\gamma_{\mu_2} \dots \gamma_{\mu_{2n}} \gamma_5 \gamma_{\mu_1}] + \mathcal{O}(\epsilon), \quad n \ge 3.
\end{equation}
Since the difference is evanescent, traces that multiply divergent integrals can give rise to finite, inconsistent results if not handled with care. As mentioned in the introduction, there are reading point prescriptions, based on conventionally keeping the same reading point in all traces \cite{Korner:1991sx, Kreimer:1993bh}, but there is no proof that this works in general. In particular, it has been shown that at three-loop level this no longer holds \cite{Chen:2023ulo}. 

On the contrary, the \textit{Breitenlohner-Maison-t'Hooft-Veltman} (BMHV) scheme has been established to be algebraically consistent at all orders. Relaxing the first property (i), the $\gamma$ matrices are split into their four-dimensional ($\gamma_{\bar{\mu}}$) and ($D-4$)-dimensional ($\gamma_{\hat{\mu}}$) components, and satisfy different anticommutation properties:
\begin{equation}
    \{\gamma_{\bar{\mu}},\gamma_5\}=0, \qquad [\gamma_{\hat{\mu}},\gamma_5]=0.
\end{equation}
This is enough to ensure algebraic consistency. However, the regularized kinetic term of fermions does not preserve chiral symmetries in this scheme, because the evanescent component mediates left-right interactions. This breaking in the regularized classical action propagates into the quantum effective action at loop order, consequently obtaining a theory that does not respect the original chiral symmetries at quantum level. If the symmetry is gauged, the consistency of the theory gets spoiled by the regularization scheme. Nevertheless, since the breaking is unphysical, the symmetry can be restored by the addition of the appropriate local counterterms. 

In the case of gauge theories, their quantization inevitably breaks the original symmetry, leaving the action invariant under the remnant BRST symmetry. When using BMHV, one option is therefore to compute the aforementioned counterterms to restore the BRST symmetry broken by the regularization. These counterterms have already been computed for some gauge theories \cite{Stockinger:2023ndm,Belusca-Maito:2020ala,Belusca-Maito:2021lnk}.
Another option is to use the background field method (BFM). This consists in splitting the fields into a classical, background part and a quantum fluctuation, and only fix the gauge for the latter. This leaves the quantum action invariant under background gauge symmetry. The counterterms in BFM have been computed for renormalizable theories in \cite{Cornella:2022hkc,OlgosoRuiz:2024dzq}, but they have yet to be computed for EFTs and SMEFT in particular.

In practice, the difference between the two schemes lies in the following anticommutator:
\begin{equation}
    \{\gamma_\mu,\gamma_5\} =
    \begin{cases*}
      0 & \text{for NDR}, \\
      2\gamma_{\hat{\mu}}\gamma_5 & \text{for BMHV},
    \end{cases*}
\end{equation}
which is, as it should, an evanescent operator. At one loop, this implies that only the finite parts of divergent loop amplitudes (with either closed or open fermion lines) or of amplitudes divergent in intermediate steps are different in the two schemes, and only in chiral theories. 
Indeed, for vector-like theories, NDR is a perfectly well-defined scheme and is  equivalent to BMHV. 
Moreover, it implies that the (finite) differences are local and, as such, they can be absorbed in the coefficient of local operators.  

Thus, we can define a shift $\Delta \coeff{i}{}$ in the Wilson coefficients of the (BMHV) SMEFT basis $\Delta S_{\mathrm{SMEFT}}^{(1)}=\int dx\, \Delta \coeff{i}{}\mathcal{O}_i$ such that:
\begin{equation}
\label{eq:defdifference}
    \Gamma^{(1)}_{\mathrm{NDR}}\equiv \Gamma^{(1)}_{\mathrm{BMHV}}+\Delta S^{(1)}_{\mathrm{SMEFT}},
\end{equation}
where the superscript $^{(1)}$ indicates one-loop order and $\Gamma$ is the 1PI effective action of the standard theory. This shift can be computed explicitly from Eq.~\eqref{eq:defdifference}, by matching the difference $ \Gamma^{(1)}_{\mathrm{NDR}}- \Gamma^{(1)}_{\mathrm{BMHV}}$ onto the SMEFT basis. Moreover, since this difference originates from evanescent pieces hitting the poles of divergent amplitudes, it is enough to compute the hard region of the quantum effective action to extract the UV poles. 

The differences in the finite pieces can lead to different bounds on the Wilson coefficients in both schemes in a bottom-up approach, see \cite{inprep}. At higher-loop orders the differences no longer need to only appear in the finite pieces. As for instance shown in \cite{DiNoi:2023ygk} in the context of SMEFT there can be also an apparent difference in the two-loop renormalisation group equations (RGEs), that though is re-absorbed by other Wilson coefficients. The scheme-dependence of the anomalous dimension matrix arising at two-loop order was firstly studied in the context of $b\to s$ transitions \cite{Ciuchini:1993ks,Ciuchini:1993fk}. In a bottom up approach it numerically impacts though the RGE running \cite{DiNoi:2023onw, DiNoi:2024ajj}. 

Of course, to perform this computation consistently, both sides of Eq. (\ref{eq:defdifference}) need to be gauge invariant, which in chiral gauge theories implies the addition, as discussed above, of symmetry-restoring counterterms in the BMHV prescription. In the case of the SMEFT, these counterterms are not yet computed, so we decided to perform this translation in the limit $g_1,g_2\rightarrow 0$, in which the SM gauge group reduces to $\mathrm{SU}(3)$, which is vector-like. 

Besides, we \textit{define} the regularized action in BMHV prescription with purely four-dimensional $\gamma$ matrices, employing the substitutions 
\begin{equation}
\gamma_{\bar{\mu}}P_{L/R}\rightarrow P_{R/L}\gamma_\mu P_{L/R} \quad \text{and} \quad \sigma_{\mu\nu}\rightarrow \sigma_{\bar{\mu}\bar{\nu}}
\label{eq:gammamuPLR}
\end{equation}
except in the fermionic kinetic term, whose treatment we will specify later. This choice, following Refs.~\cite{Ciuchini:1993vr, Belusca-Maito:2023wah, Cornella:2022hkc}, is purely a definition of a renormalization scheme, and any other choice could be put in this form by a finite renormalization (see more details in Section \ref{sec:matching}).

Finally, even though in this limit of vanishing electroweak couplings the gauge symmetry is not chiral, the SMEFT Lagrangian is, by construction, invariant under a global $\mathrm{SU}(2)\times\mathrm{U}(1)$ symmetry. This symmetry is violated by the regularization procedure, so $\Gamma^{(1)}_{\mathrm{BMHV}}$ will not respect this global invariance. Since the right hand side in Eq. (\ref{eq:defdifference}) is invariant, this leads to an inconsistency. One has two choices to fix this issue. The first one would be to accept the breaking of this global symmetry, which was only accidental in the limit we are considering, and add a basis of $\mathrm{SU}(2)\times\mathrm{U}(1)$-violating operators to $\Delta S^{(1)}_{\mathrm{SMEFT}}$ to match the offending terms. Alternatively, as we decided to do in this work, one can maintain the symmetry by adding the appropriate counterterms to the BMHV effective action. 

We opted for this second option because these counterterms are a subset of those which would be added when restoring the full, gauge $\mathrm{SU}(2)\times\mathrm{U}(1)$ symmetry, that would be needed to perform this translation in the full SMEFT. The global-symmetry-restoring counterterms are, however, particularly simple to compute in our case. To this end, we follow the technique developed in \cite{OlgosoRuiz:2024dzq}, introducing a spurion field $\Omega$ to recover the chiral symmetry. Defining a fermionic multiplet $f=(u,d,\nu,e)^T$ we can write a gauge invariant kinetic term for the SM as:
\begin{equation}
\label{eq:kin_term_ferm}\mathcal{L}_{\mathrm{kin}}\supset\bar{f}i\overline{\slashed{\partial}}f + \bar{f_L}i\Omega\hat{\slashed{\partial}} f_R +\bar{f_R}i\Omega^\dagger\hat{\slashed{\partial}} f_L
\end{equation}
with $\Omega$ an unitary matrix transforming as $\Omega\rightarrow [U^{(L)}_{\mathrm{SU(2)}}\otimes U^{(L)}_{Y}] \,\Omega\, [U^{(R)\dagger}_{Y}]$, and:
\begin{equation}
    U_Y^{(L/R)}=e^{i\theta T^{L}_{Y}},\qquad
    T^{L}_{Y}=\begin{pmatrix}
        \frac16{\mathbb 1}&&&\\
        &\frac16{\mathbb 1}&&\\
        &&-\frac12&\\
        &&&-\frac12
    \end{pmatrix},\qquad
    T^{R}_{Y}=\begin{pmatrix}
        \frac23{\mathbb 1}&&&\\
        &-\frac13{\mathbb 1}&&\\
        &&0&\\
        &&&-1
    \end{pmatrix},
\end{equation}
\begin{equation}
    U_{\mathrm{SU(2)}}^{(L)}=e^{i\theta_a T^a},\qquad
    T^{a}=\begin{pmatrix}
       \frac12\sigma^a{\mathbb 1}&&\\
        &\frac12\sigma^a
    \end{pmatrix},
\end{equation}
with $\theta_{(a)}$ the parameter of the global transformation and $\mathbb{1}$ the $3\times3$ identity matrix in colour space. Since the SM gauge group does not mix quark and leptons, we can further parametrize $\Omega$ as:
\begin{equation}
    \Omega\equiv\begin{pmatrix}
        \Omega_q & \mathbb 0\\
        \mathbb 0 & \Omega_\ell
    \end{pmatrix},
\end{equation}
with $\Omega_{q(\ell)}$ a $2\times 2$ unitary matrix for the quark (lepton) sector. Using a quark ($q$) and lepton ($\ell$) doublet we can rewrite Eq. (\ref{eq:kin_term_ferm}) as:
\begin{align}
\label{eq:kin_term_ferm_bidoublet}
\mathcal{L}_{\mathrm{kin}}\supset&\, \overline{q}i\overline{\slashed{\partial}} q +\overline{\ell}i\overline{\slashed{\partial}} \ell +\Big[\overline{q}_L\Omega_q i\hat{\slashed{\partial}} q_R +\overline{\ell}_L\Omega_{\ell} i\hat{\slashed{\partial}} \ell_R +\mathrm{h.c.}\Big].
\end{align}

The quantum effective action defined with a non-dynamical $\Omega$ is gauge invariant by construction and, in general, can be splitted in a part explicitly containing $\Omega$ ($\Gamma_{\Omega}$) and a part independent of $\Omega$ ($\Gamma_\slashed{\Omega}$). As shown in \cite{OlgosoRuiz:2024dzq}, adding a set of local counterterms to cancel $\Gamma_{\Omega}$ restores the gauge invariance of the original theory, which is recovered by setting $\Omega\rightarrow 1$. In general, identifying these counterterms is only straightforward if one uses a space-time dependent $\Omega$ that allows to reconstruct its covariant derivatives since some of these counterterms contain terms that actually do not depend on $\Omega$. However, in the case of a global chiral symmetry, one can choose $\Omega$ to be a constant so that there are no counterterms with $D_\mu \Omega$ and they are exactly equivalent to $\Gamma_\Omega$. The only $\Omega$-dependence comes through the definition of the fermion propagators from Eq. (\ref{eq:kin_term_ferm_bidoublet}). Thus, cancelling $\Gamma_\Omega$ equals to send $\Omega\rightarrow 0$ after imposing $\Omega^\dagger\Omega=\mathbb 1$ (see \cite{OlgosoRuiz:2024dzq} for further details on this discussion).

Therefore, taking all these considerations into account, we compute the difference between the two schemes using the formula:
\begin{equation}
    \Delta S^{(1)}_{\mathrm{SMEFT}}=\Gamma^{(1)}_{\mathrm{NDR}}|^{(h)}-\Gamma^{(1),\,\slashed{\Omega}}_{\mathrm{BMHV}}|^{(h)},
\end{equation}
where the superscript $(h)$ denotes the hard region of the quantum effective action, in which the loop momentum is much larger than any other scale.

\section{Examples} \label{sec:examples}
In order to clarify how we perform the calculation in practice, we discuss in detail the specific example of two different contributions to the Yukawa corrections. The first contribution is the Yukawa self-correction of $\mathcal{O}(y^3)$ computed in the Green's basis, while the second example, the four-quark contributions, is computed on-shell to match the results of \cite{DiNoi:2023ygk}. 

Given that the spurion $\Omega$ intertwines the $\mathrm{SU(2)}$ quark components, it is convenient to work with a quark doublet for both chiralities. For instance, the SM Yukawa Lagrangian can be written in the following form:
\begin{align}
\label{eq:lag_yuk}
    -\mathcal{L}_{\mathrm{Yuk}}&=\overline{q}_i [ H_a y_d \delta_{2,j}\delta_{ia} + H_a^\dagger \varepsilon_{ia}\delta_{1,j} y_u ]q_{R,j} \nonumber\\
    &+\overline{q}_i [ H_a \varepsilon_{a,j}\delta_{1,i} y_u^\dagger + H_a^\dagger \delta_{aj} \delta_{2,i} y_d^\dagger  ]q_{L,j},
\end{align}
where the indices indicate the $\mathrm{SU(2)}$ components and flavour and color indices are omitted. 

\subsection{Yukawa self-correction}

\begin{figure}[h]
    \centering
        \begin{subfigure}{.32\textwidth}
        \centering
        \begin{tikzpicture}[baseline=(y)]
            \begin{feynman}
                \vertex (y) [scale = \sizedot, dot, black] {};
                \vertex (h) [left = of y ] {$H$};
                \vertex (u1) [above right =45 pt of y] {$q$};
                \vertex (u2) [below right = 45 pt of y] {$q$};

                \vertex (y1) [scale=\sizedot,dot,above right =20 pt of y] {};
                \vertex (y2) [scale=\sizedot,dot,below right =20 pt of y] {};
                \diagram* {
                    (h) -- [charged scalar] (y), 
                    (u1) -- [fermion] (y1) -- [fermion] (y) -- [fermion] (y2) -- [fermion] (u2), 
                    (y2) --[charged scalar] (y1),
                };
            \end{feynman}
        \end{tikzpicture}
        \subcaption{}\label{fig:diagEX1SM1L}
    \end{subfigure}
 \begin{subfigure}{.32\textwidth}
        \centering
        \begin{tikzpicture}[baseline=(y)]
            \begin{feynman}
                \vertex (y) [scale = \sizedot, dot, black] {};
                \vertex (h) [left = of y ] {$H$};
                \vertex (u1) [above right =45 pt of y] {$q$};
                \vertex (u2) [below right = 45 pt of y] {$q$};

                \vertex (y1) [scale=\sizedot,dot,above right =20 pt of y] {};
                \vertex (y2) [scale=\sizedot,dot,below right =20 pt of y] {};
                \diagram* {
                    (h) -- [charged scalar] (y), 
                    (u1) -- [fermion] (y1) -- [fermion] (y) -- [fermion] (y2) -- [fermion] (u2), 
                    (y1) --[charged scalar] (y2),
                };
            \end{feynman}
        \end{tikzpicture}
        \subcaption{}\label{fig:diagEX1SM1L}
    \end{subfigure}
       \begin{subfigure}{.32\textwidth}
        \centering
        \begin{tikzpicture}[baseline=(4F)]
            \begin{feynman}
                \vertex (4F) [scale = \sizesqdot, black, square dot] {};
                \vertex (y) [scale = \sizedot, dot, black, left =30 pt of 4F] {};
                \vertex (h) [left = of y ] {$H$};
                \vertex (u1) [above right = of 4F] {$q$};
                \vertex (u2) [below right = of 4F] {$q$};
                \diagram* {
                    (h) -- [charged scalar] (y), 
                    (u1) -- [fermion] (4F) -- [fermion] (u2), 
                    (y) -- [fermion, half right] (4F) -- [fermion, half right] (y),
                };
            \end{feynman}
        \end{tikzpicture}
        \subcaption{} \label{fig:diagEX14t}
    \end{subfigure}
    \caption{Feynman diagrams contributing to the process $H_3 \to \bar{q}_1q_2$ within the SM (a,b) and with four-quark SMEFT operators (c). }
    \label{fig:diagEX1}
\end{figure}
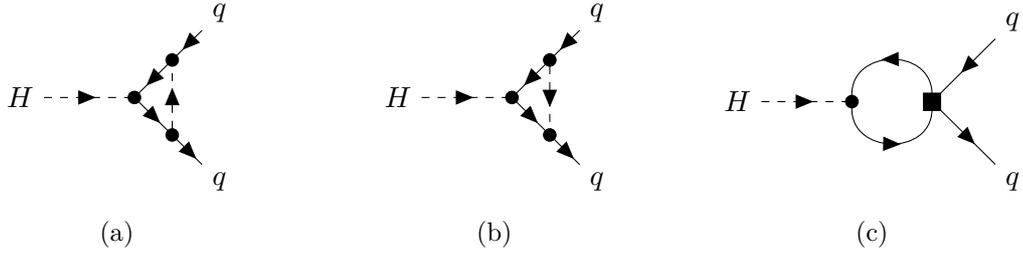

Let us start by considering the one-loop contribution arising in the SM in Fig.~\ref{fig:diagEX1} 
to the process $H_3 \to \bar{q}_1 q_2$. 
To compute the shift to $y_{u,d}$ we compute the one-loop amplitude in both BMHV and NDR schemes, extract their hard region and take the difference. We take all momenta to be incoming. 
The results for both diagrams are the following: 
\begin{align}
i\mathcal{M}_{(a)}^{\mathrm{(NDR-BMHV)}} &= \Big\{\overline{v}_1 \PL u_2 \,\delta_{2,l_1} [\varepsilon \Omega ]_{l2,1} [\varepsilon \Omega]_{l_3,1} [y_d^\dagger y_u^\dagger y_u^\dagger] \nonumber\\
&+ \overline{v}_1 \PR u_2 \Big[ \delta_{l_2,2}\delta_{l_1,l_3} [y_u y_u^\dagger y_d] + [\Omega \varepsilon]_{2,l_1}\Omega_{1,l_3}\delta_{l_2,2}[y_u y_d y_d]\Big]\Big\}\nonumber\\
& \times \int \frac{d^Dk}{(2\pi)^D}\left(\frac{4-D}{D}\right)\frac{1}{k^4},
\end{align}
\begin{align}
    i\mathcal{M}_{(b)}^{\mathrm{(NDR-BMHV)}}&=\Big\{ \overline{v}_1\PL u_2 \Big[ \delta_{l_1,1}\varepsilon_{l_2,l_3} [y_u^\dagger y_d y_d^\dagger] - \delta_{l1,1} [\varepsilon \Omega]_{l_3,2} [\varepsilon \Omega]_{l_2,1}[y_u^\dagger y_u^\dagger y_d^\dagger]\Big]\nonumber\\
    &+\overline{v}_1 \PR u_2 \Big[ -\delta_{l_2,1} \Omega^\dagger_{2,l_3} [\Omega^\dagger \varepsilon ]_{2,l_1}[y_d y_d y_u]  \Big]\Big\}\nonumber\\
    & \times \int \frac{d^Dk}{(2\pi)^D}\left(\frac{4-D}{D}\right)\frac{1}{k^4},
\end{align}
where $l_i$ denotes the $\mathrm{SU(2)}$ component of the $i$-th particle and flavor and color indices are omitted. We have set to zero the external momenta, since we are interested in extracting the correction for the Yukawa interactions. As expected, the result is proportional to $D-4$ and, as such, will produce a finite, local contribution when multiplying the pole of the integral, which can be extracted from its hard-region using $\int 1/k^4\rightarrow 1/\epsilon_{\mathrm{UV}}$. We left explicit the $\Omega$ dependence to see how the ``na\"ive'' calculation in BMHV, i.e. setting $\Omega\rightarrow 1$, produces terms that cannot be absorbed by the $\mathrm{SU(2)}$-symmetric operators in Eq. (\ref{eq:lag_yuk}), which need for the inclusion of $\mathrm{SU(2)}$-violating structures. In our approach, we send $\Omega\rightarrow 0$ and, equating to the tree-level amplitude from Eq. (\ref{eq:lag_yuk}), we obtain:
\begin{equation}
    \Delta y_d = -\frac{y_u y_u^\dagger y_d}{32\pi^2}, \hspace*{1cm}
    \Delta y_u^\dagger = -\frac{y_u^\dagger y_d y_d^\dagger}{32\pi^2}.
\end{equation}
\subsection{Four-quark Operators}

We now consider the contribution to the Yukawa coupling from the four-quark operators $\op{qu(1,8)}{}$, see Fig.~\ref{fig:diagEX14t}.  
For simplicity, we focus on the contribution to the up-quark Yukawa coupling.

The Feynman rules for the SMEFT operators are given in Ref.~\cite{Dedes:2017zog}. 
As already detailed, the difference between the two schemes can arise only from the rational terms, which we report here: 
\begin{equation} \label{eq:Ytcorrections}
\begin{aligned}
i\mathcal{M}_{(c)}^{\mathrm{(NDR-BMHV)}} &= i\left( \coeff{qu(1)}{prst} + C_F\, \coeff{qu(8)}{prst} \right) \, \dfrac{ m_H^2 \yuk{u}^{\dagger,tp}}{16 \pi ^2 }  \\ & \times 
\left[ 
\frac{5}{3}  \left(\bar{v}_1\PR  u_2 \right) \delta_{l_1,1} \varepsilon_{l_3,l_2}    
-  \Omega^\dagger_{l_1,1}  \varepsilon_{l_3,l'}\Omega ^\dagger_{l',l_2} \, \dfrac{2}{3} \,  \left(\bar{v}_1\PL  u_2 \right)\right].    
\end{aligned}
\end{equation}
To avoid ambiguities, we have reintroduced explicitly the flavour indices. We call $\coeff{qu(1)/(8)}{}$
the Wilson coefficient of the operators $\mathcal{O}_{qu(1)/(8)}$.
In the previous expression $C_F=4/3$ is the quadratic Casimir invariant for $SU(3)$. 
We stress that this result has been computed on-shell, setting the Higgs momentum $q^2 = m_H^2$. 

As in the previous case, the BMHV result features a term which explicitly breaks the chiral symmetry. 
In order to restore the broken symmetries symmetry-restoring finite counterterms are added. Due to the addition of the spurion field, a shortcut for the inclusion of the counterterms is simply setting $\Omega \to 0$, after using the unitarity property $\Omega\Omega^\dagger = \Omega^\dagger \Omega =\mathbb{1}$ at intermediate stages. 

Dropping the terms proportional to $\Omega$ in Eq.~\eqref{eq:Ytcorrections} yields a result which is completely proportional to $\PR$, as expected from the chiral symmetry. 
We obtain:
\begin{equation}
\Delta\yuk{u}^{\dagger,rs}  = +  \frac{5}{3} \left( \coeff{qu(1)}{prst} + C_F\, \coeff{qu(8)}{prst} \right) \, \dfrac{  m_H^2  \yuk{u}^{\dagger,tp}}{16 \pi ^2 } .
\end{equation}

An alternative approach would be using na\"ively the fermion propagator also in BMHV, which means setting $\Omega=1$.
The result violates manifestly the chiral symmetry. However, it is possible to focus on the interactions with the physical Higgs boson only, which would mean considering only the term proportional to the identity in Dirac space in Eq.~\eqref{eq:Ytcorrections}. This computation is more intuitive in the broken phase.  
This choice reproduces the result in Ref.~\cite{DiNoi:2023ygk},
\begin{equation}
\label{eq:shift_yuk_oldpaper}
\Delta\yuk{u}^{\dagger,rs}  =
  \left( \coeff{qu(1)}{prst} + C_F\, \coeff{qu(8)}{prst} \right) \, \dfrac{ m_H^2  \yuk{u}^{\dagger,tp}}{16 \pi ^2 },
\end{equation}
up to a minus sign, once we send $m_H^2\rightarrow -\lambda v^2$ to match the conventions. 
Indeed, the results of this paper represent the scheme-dependent terms projected to the Warsaw basis. In order to achieve a scheme-independent result, as in Ref.~\cite{DiNoi:2023ygk}, the contributions computed here must be subtracted accordingly, defining a renormalization scheme and explaining the difference in sign.

\section{Results \label{sec:results}}
In this section, we will discuss how we obtained our results. Due to their length, they will be provided in an ancillary file.

\subsection{Implementation}

Due to the large number of diagrams to compute for translating the complete SMEFT up to dimension six, the task is prone to automation. Given that we have to compute the hard region of a one-loop effective action, we can make use of \texttt{Matchmakereft} \cite{Carmona:2021xtq}, an automated tool for matching calculations in effective field theories. Since \texttt{Matchmakereft} is capable (in a version that will be made public in the near future) of doing loop calculations in the BMHV scheme, it is well-suited for our task.

Since \texttt{Matchmakereft} performs calculations off-shell, in a first step we defined a model for the SMEFT Green's basis, expressed only in terms of $\mathrm{SU(2)}$ doublets and with $g_1, g_2\rightarrow 0$. This model used at tree-level will act as a basis for $\Delta S_{\mathrm{SMEFT}}$ absorbing the difference between schemes.

To compute the one-loop effective action, we define
two models, one for NDR and one for BMHV, with only physical operators.
The operators in the BMHV scheme were defined strictly four-dimensional, as in Eq.~\eqref{eq:gammamuPLR}. The propagators are also modified to include the $\Omega$ dependence as in Eq.~\eqref{eq:kin_term_ferm}.

Then, we computed the pertinent one-loop amplitudes, both in BMHV and NDR, took the difference, and extracted the finite part of its hard-region. This just required minimal modification of the original \texttt{Matchmakereft} set-up that extracts the UV poles for RGE calculations. 
After imposing th unitarity of $\Omega$, we sent $\Omega\rightarrow 0$ to restore the global $\mathrm{SU(2)\otimes U(1)}$ symmetry, and matched to the tree-level $\Delta S_{\mathrm{SMEFT}}$. Finally, the result is canonically normalized and reduced to the SMEFT physical (Warsaw) basis.

We manually cross-checked several shifts involving different classes of operators making use of \texttt{FeynCalc} \cite{Mertig:1990an, Shtabovenko:2023idz}, agreeing with the automatized computation.

We provide our results as a \texttt{Mathematica} replacement list, where the rule for each coefficient returns its corresponding shift in the physical basis. The sign of this shift corresponds to the difference NDR - BMHV, i.e. it is the shift to be added to the BMHV result to obtain the NDR one.
We follow the conventions from \texttt{Matchmakereft} for the SM Lagrangian and nomenclature of Wilson Coefficients, denoted as \texttt{alphaOxx} for the operator \texttt{Oxx}. We refer to Ref.~\cite{Carmona:2021xtq} for further details.

\subsection{About ambiguous traces in NDR}

As discussed in Sec.~\ref{sec:framework}, traces of $\gamma_5$ with six or more $\gamma$ matrices are ambiguous. These traces can in principle appear in our calculations so we will use a reading point prescription to ensure the coherence and reproducibility of our results. Given that the ambiguity is proportional to the Levi-Civita tensor, the potential ambiguities will be in operators of the Green's basis with the schematic form $\overline{q}\slashed{D}q \widetilde{G}_{\mu\nu}$, $ \overline{q}q \widetilde{G}_{\mu\nu} H$, $H^\dagger H G \widetilde{G}$, $GG\widetilde{G}$, fixed by the amplitudes $\overline{q} q \to G $, $\overline{q} q \to G H$, $H^\dagger H \to G G$ and $G G \to G$, respectively. 
The requirement of a closed fermion loop and the number of $\gamma$ matrices greatly restrict the possible contributions to the ambiguity. The amplitude $\overline{q} q \to G $ can only be mediated by a four-fermion operator, and it does not contain enough $\gamma$ matrices in any case. The amplitude $\overline{q} q \to G H$ can, again, only have a closed fermion loop with a four fermion operator. The only insertion that gives more than six $\gamma$ matrices is the operator $\mathcal{O}_{lequ}^{(3)}$ that, in our limit $g_{1,2}\rightarrow 0$, cannot contribute to this process. The amplitude $G G \to G$ could only give an ambiguity through the usual SM triangle diagram, that vanishes due to anomaly cancellation. No SMEFT operators enter here.

Finally, the amplitude $H^\dagger H \to G G$ is the only one that can generate ambiguous contributions, absorbed by the coefficient $\alpha_{H\tilde{G}}$. We performed the calculation implementing a fixed reading point procedure in \texttt{Matchmakereft}, so that all traces are started from the same vertex.
We found that the ambiguous result is purely imaginary, so imposing the hermiticity of the action one can actually fix the ambiguity in this case given that $\alpha_{H\tilde{G}}$ is real. Nevertheless, for completeness, we decided to include this (unphysical) result for $\alpha_{H\tilde{G}}$ parametrized by a coefficient $\xi_{\mathrm{RP}}=\pm 1$, which indicates whether the traces should be read starting from the vertex with $H$ or $H^\dagger$, respectively.

\section{Scheme differences in matching calculations \label{sec:matching}}
A hidden assumption that we have made when writing Eq.~\eqref{eq:defdifference} is that the one-loop SMEFT Wilson Coefficients, that enter $\Gamma^{(1)}_{\mathrm{EFT}}$ at tree level, are the same in both schemes. Indeed, expanding Eq. (\ref{eq:defdifference}):
\begin{align}
    \Gamma_{\mathrm{BMHV}}^{(1)}&=S^{(0)}[\coeff{\text{BMHV}}{(1)}]+\Gamma_{\mathrm{BMHV}}^{(1)}[\coeff{}{(0)}]=\nonumber\\
    &=S^{(0)}[\coeff{\text{NDR}}{(1)}]+\Gamma_{\mathrm{NDR}}^{(1)}[\coeff{}{(0)}]+\Delta S_{\mathrm{SMEFT}},
\end{align}
where $S^{(0)}$ is the tree-level action and $\Gamma^{(1)}[\coeff{}{(0)}]$ denotes the 1-particle-irreducible diagrams with insertions of tree-level coefficients. 

This equation, obviously, only holds if $\coeff{\text{NDR}}{(1)}=\coeff{\text{BMHV}}{(1)}$, with $\coeff{}{(1)}$ the one-loop order matching contribution to the Wilson Coefficients, which is implicitly imposed when doing bottom-up calculations in the EFT. 
 However, matching calculations are of course susceptible of receiving scheme differences at loop order, for the same reasons stated in Sec. \ref{sec:framework}. This would reintroduce a difference between schemes that can be, again, restored by the addition of appropriate local, finite shifts. 

In order to perform the matching, it is very useful to use the method of expansion by regions. This fixes the one-loop matching coefficients by imposing:
\begin{align}
    S^{(0)}_{\mathrm{EFT}}[\coeff{}{(1)}]=\Gamma_{\mathrm{UV}}^{(1),\mathrm{1lPI}}|^{(h)},
\end{align}
where $\Gamma^{\mathrm{1lPI}}|^{(h)}$ denotes the hard-region of the one-light-particle-irreducible effective action.
The scheme dependence manifests in the following way. The finite part of the hard-region UV effective action can be written as:
\begin{align}
\label{eq:hardregion_matching_scheme}
    \Gamma_{\mathrm{UV}}|^{(h)}=\overline{\Gamma}^{(h)} + \Gamma^{\epsilon}\Big(\frac{A}{\epsilon_{ \mathrm{\scriptscriptstyle{UV}}}}+\frac{B}{\epsilon_{ \mathrm{\scriptscriptstyle{IR}}}}\Big).
\end{align}
$\overline{\Gamma}$ is the scheme independent part and $\Gamma^{\epsilon}$,  which is $\mathcal{O}(\epsilon)$, is different in the two schemes. $\epsilon_{ \mathrm{\scriptscriptstyle{UV}}}$ are the UV poles of the UV theory, while $\epsilon_{ \mathrm{\scriptscriptstyle{IR}}}$ are spurious IR divergences that arise from the hard region expansion.

If we insert this in the EFT effective action, we get:
\begin{align}
    \Gamma_{\mathrm{EFT}}^{(1)}
    &=S^{(0)}[\coeff{}{(1)}]+\overline{\Gamma}_{\mathrm{EFT}}+\Gamma^{\epsilon}_{\mathrm{EFT}}\Big(\frac{B'}{\epsilon_{ \mathrm{\scriptscriptstyle{UV}}}}\Big)=\nonumber\\
    &=\overline{\Gamma}_{\mathrm{UV}}^{(h)} + \Gamma^{\epsilon}_{\mathrm{UV}}\Big(\frac{A}{\epsilon_{ \mathrm{\scriptscriptstyle{UV}}}}+\frac{B}{\epsilon_{ \mathrm{\scriptscriptstyle{IR}}}}\Big)+\overline{\Gamma}_{\mathrm{EFT}}+\Gamma^{\epsilon}_{\mathrm{EFT}}\Big(\frac{B'}{\epsilon_{ \mathrm{\scriptscriptstyle{UV}}}}\Big)=\nonumber\\    &=\overline{\Gamma}_{\mathrm{UV}}^{(h)}+\overline{\Gamma}_{\mathrm{EFT}}+ \Gamma^{\epsilon}_{\mathrm{UV}}\Big(\frac{A}{\epsilon_{ \mathrm{\scriptscriptstyle{UV}}}}\Big)\equiv \overline{\Gamma}_{\mathrm{UV}}+ \Gamma^{\epsilon}_{\mathrm{UV}}\Big(\frac{A}{\epsilon_{ \mathrm{\scriptscriptstyle{UV}}}}\Big),
\end{align}
 where $\Gamma_{\mathrm{EFT}}^{(1)}$ can be either $\Gamma_{\mathrm{NDR}}^{(1)}$ or $\Gamma_{\mathrm{BMHV}}^{(1)}$. Moreover, in the second line we have inserted Eq. (\ref{eq:hardregion_matching_scheme}), and in the third line we have used that $\Gamma^\epsilon_{\mathrm{UV}}B=-\Gamma^\epsilon_{\mathrm{EFT}}B'$ and $\overline{\Gamma}^{(h)}_{\mathrm{UV}}+\overline{\Gamma}_{\mathrm{EFT}}=\overline{\Gamma}_{\mathrm{UV}}$ by construction.\footnote{We refer to \cite{Manohar:2018aog} Sec.~5 explaining nicely this point.} \\ 

We can extract several lessons from this result. First, we see that the scheme dependence in the EFT cancels in the matching, as it should, so that $\Gamma_{\mathrm{EFT}}$ reproduces $\Gamma_{\mathrm{UV}}$ (of course, in some low-energy limit). Second, we see that the UV effective action can obviously also be translated between schemes by the addition of  $\Delta S_{\mathrm{UV}}\equiv \left( \Gamma^{\epsilon \mathrm{(NDR)}}_{\mathrm{UV}} -\Gamma^{\epsilon \mathrm{(BMHV)}}_{\mathrm{UV}}\right)\Big(\frac{A}{\epsilon_{ \mathrm{\scriptscriptstyle{UV}}}}\Big)$. Therefore, in this case the remaining scheme dependence of the matching coefficients, fixed by $ \Gamma_{\mathrm{UV}}|^{(h)}$, is encoded in $\Gamma^\epsilon_{\mathrm{UV}}B$, that is precisely given by the quantity $\Gamma^\epsilon_{\mathrm{EFT}}B'$ that we computed in both schemes for SMEFT.

Finally, as a consequence, notice that for amplitudes that are convergent in the UV, i.e. $A=0$, the scheme dependence in the matching is fixed by the one loop one-particle-irreducible effective action. Consequently, they cancel to give a scheme independent effective action that reproduces $\Gamma_{\mathrm{UV}}$ which is scheme independent by assumption, since the dependence arises from divergences.

To better exemplify this result we performed the complete one-loop matching of the SM extended by a heavy vector-like fermion $\Psi\sim(3,1,2/3)$ onto the SMEFT, both in the NDR and BMHV schemes. Again, to avoid the inclusion of gauge-restoring counterterms in the BMHV calculation we take the limit $g_{1,2}\rightarrow 0$ (nevertheless, we employ the same technique described above to restore the global $\mathrm{SU(2)}\otimes\mathrm{U(1)}$ symmetry). The Lagrangian for the UV model is the following:
\begin{equation}
\label{eq:lag_matching_example}
\mathcal{L} = \mathcal{L}_{\mathrm{SM}} + \overline{\Psi}\mathrm{i}\slashed{D}\Psi - M \overline{\Psi}\Psi + \Big[ y_T \overline{q}_L \Psi \varepsilon H^{\dagger} + \mathrm{h.c.}\Big],
\end{equation}
where we omitted flavor and gauge indices.
We performed the matching using \texttt{Matchmakereft}, obtaining the results that we give as an ancillary file.
All calculations are performed off-shell in $D$ dimensions, so the result is canonically normalized and subsequently reduced to the Warsaw basis. For simplicity, we only report the results in the physical basis.
To the best of our knowledge, it is the first example of an automatic complete one-loop matching calculation in the BMHV scheme with symmetry restoring counterterms.

When comparing matching results with the scheme-translating shifts that we have computed, one has to take into account that our starting point was a physical, four-dimensional EFT. Integrating-out heavy particles gives a $D$-dimensional Green's basis that has to be reduced to the former one for a correct comparison. In this reduction, one has to take special care of tree-level evanescent structures, whose effect in loops can be incorporated by a (scheme-dependent) shift in the coefficients in the physical basis \cite{Fuentes-Martin:2022vvu,Aebischer:2022aze,Aebischer:2022rxf}.

Integrating Eq. (\ref{eq:lag_matching_example}) at tree-level we have:
\begin{align}
    \mathcal{L}_{\mathrm{EFT}}^{(0)}=&-\frac{1}{2}y_T (H^\dagger \varepsilon \overline{q}_L) \frac{i\slashed{D}}{M^2}(q_L \varepsilon H)+\mathcal{O}(M^{-4})\nonumber\\
    =&\phantom{+\,}\frac{y_T y_T^\dagger}{4M^2} \Big( H^\dagger H \overline{q}_L i \overset{\leftrightarrow}{\slashed{D}} q_L  - H^\dagger \sigma^I H \overline{q}_L \sigma^I i \overset{\leftrightarrow}{\slashed{D}} q_L\Big) +\nonumber\\
    &+\frac{y_T y_T^\dagger}{4M^2} \Big( H^\dagger\overset{\leftrightarrow}{D^\mu} H \overline{q}_L i \gamma_\mu q_L  - H^\dagger \sigma^I \overset{\leftrightarrow}{D^\mu} H \overline{q}_L \sigma^I i \gamma_\mu q_L\Big)\nonumber\\
    \equiv& \,\beta_{Hq}^{' (1)}\mathcal{R}_{Hq}^{' (1)} +\beta_{Hq}^{' (3)}\mathcal{R}_{Hq}^{' (3)}  +\alpha_{Hq}^{(1)}\mathcal{O}_{Hq}^{(1)} +\alpha_{Hq}^{(3)}\mathcal{O}_{Hq}^{(3)},
\end{align}
where here we use the notation $\alpha(\beta)$ to distinguish between the coefficients of physical (redundant) operators, following Ref.~\cite{Carmona:2021xtq}.

We do not generate any evanescent structure directly. In order to reduce to the physical basis,  equations of motion need to be applied. Not having to use Fierz relations or any four-dimensional identity, the reduction in the NDR scheme does not generate any evanescent structure. However, deriving the equation of motion in the BMHV scheme from Eq. (\ref{eq:kin_term_ferm_bidoublet}) reads:
\begin{align}
    \overline{\slashed{D}}q=-\Omega_q \hat{\slashed{\partial}}q_R-\Omega_q^\dagger \hat{\slashed{\partial}}q_L + \text{non evanescent terms}.
\end{align}
So we implicitly define the following evanescent lagrangian:
\begin{align}
\label{eq:evanescent_lag_BMHV}
    \mathcal{L}_{\mathrm{ev}}^{(0)}=-\beta_{Hq}^{' (1)} H^\dagger H \overline{q}_L i \Omega_q \hat{\slashed{\partial}} q_R-\beta_{Hq}^{' (3)} H^\dagger\sigma^I H \overline{q}_L i \Omega_q \sigma^I \hat{\slashed{\partial}} q_R
    +\mathrm{h.c.}
\end{align}
Therefore, we need to compute the shift to $c^{(1)}_{\mathrm{BMHV}}$ induced by Eq.~\eqref{eq:evanescent_lag_BMHV} to compare correctly.
Indeed, if we examine, for instance, the matching contribution to the chromomagnetic operator $\mathcal{O}_{dG}$ in the physical basis, we can see that 
\begin{align}
    (\delta \alpha_{dG})_{ij}=-\frac{g_3 (y_T)_i (y_T)^\dagger_{k}(y_d)_{kj}}{96 \pi^2 M^2},
\end{align}
where we defined $\delta \alpha_{dG}\equiv\alpha_{dG}^{\mathrm{NDR}}-\alpha_{dG}^{\mathrm{BMHV}}$. The shift to $\alpha_{dG}$ that we computed reads instead:
\begin{align}
    (\Delta \alpha_{dG})_{ij}=\frac{g_3 (y_T)_i (y_T)^\dagger_{k}(y_d)_{kj}}{192 \pi^2 M^2}.
\end{align}
However, correctly accounting for the evanescent correction stemming from Eq.~\eqref{eq:evanescent_lag_BMHV} modifies $\alpha_{dG}^{\mathrm{BMHV}}$ in such a way that:
\begin{align}
    \alpha_{dG}^{\mathrm{BMHV}}&\rightarrow \alpha_{dG}^{\mathrm{BMHV}}+\frac{(\beta^{'(1)}_{Hq}+3\,\beta^{'(3)}_{Hq})_{i k}(y_d)_{kj}}{96\pi^2}\nonumber\\
    &\Rightarrow(\delta \alpha_{dG})_{ij}=-\frac{g_3 (y_T)_i (y_T)^\dagger_{k}(y_d)_{kj}}{192 \pi^2 M^2},
\end{align}
so that the scheme differences cancel, as it should, since the UV amplitude of $\bar{q}_L d_R \to H G$ is convergent. The result presented in the ancillary file contains this evanescent correction.  We stress that this does not have to be the case in general. In particular, kinetic terms receive scheme-dependent corrections which stem from UV poles that, when normalizing canonically, propagate into several coefficients.

\section{Conclusions \label{sec:conclusion}}
Dimensional regularization is commonly used in the particle physics community to perform loop computations. In chiral theories, a prescription for the treatment of $\gamma_5$, which is purely four-dimensional, needs to be defined. The most common choice, na\"ive dimensional regularization, is widely adopted because of its simplicity, but gives ambiguous results in higher order calculations. On the other hand, the BMHV scheme is algebraically consistent, but more cumbersome to implement. 

In this work we have studied the relation between both schemes in an EFT framework, showing how EFT loop calculations can be translated between both schemes via a finite renormalization. In particular, we computed the translating shifts for the SMEFT up to dimension six, in the limit $g_1, g_2\rightarrow 0$. This simplification allows to avoid the inclusion of gauge-restoring counterterms in BMHV.
Once these gauge-restoring counterterms are computed for the SMEFT, the extension of our results is technically straightforward.

Likewise, we have shown how matching calculations introduce a scheme-dependence in the Wilson Coefficients, that could always be compensated with a finite renormalization in the same fashion, but in a model dependent way. In other words, one could translate a NDR-computed one-loop process in SMEFT to the BMHV result using our shifts, but should then perform matching in BMHV to interpret the result in specific UV scenarios.
As an example, we performed the complete one-loop matching of the SM extended with a heavy up-type quark in both schemes, again introducing symmetry-restoring counterterms, using \texttt{Matchmakereft}. We chose an example in which, after combining the matching results and the one-loop EFT computation, the scheme dependence dropped out. This though holds only true in general in the absence of UV divergencies in the UV theory. We emphasise that both schemes are, in absence of ambiguous traces, valid choices.
Any remaining differences after consistent matching can be considered as differences in the renormalization scheme, which can be fixed with finite counterterms.



At higher loop orders, where ambiguous traces are recurrent in calculations in the NDR scheme, it is crucial that we have a consistent prescription, to ensure the validity of our results. The automatization of these calculations in the BMHV scheme would be very convenient. In that sense, using our mapping avoids the implementation of BMHV in existing automatic EFT tools, making it a very efficient way of recycling already known NDR results in the BMHV scheme.

\acknowledgments
This work received funding by the INFN Iniziativa Specifica APINE,  by the Italian MUR Departments of Excellence grant 2023-2027 “Quantum Frontiers”, by the University of Padua under the 2023 STARS Grants@Unipd programme (Acronym and title of the project: HiggsPairs – Precise Theoretical Predictions for Higgs pair production at the LHC), by the European Union’s Horizon 2020 research and innovation programme under the Marie Sklodowska-Curie
grant agreement n. 101086085 – ASYMMETRY, by the PNRR CN1-Spoke 2 and by the Italian Ministry of University and Research (MUR) via the PRIN 2022 project n. 20225X52RA — MUS4GM2 funded by the European Union via the Next Generation EU package.


\bibliographystyle{JHEP}\bibliography{biblio.bib}


\end{document}